\documentclass[aps,prb,showpacs,twocolumn,superscriptaddress]{revtex4}
\usepackage[dvips]{graphicx,color}
\usepackage{amssymb}
\usepackage{xcolor,ulem}

\begin{document}
\title{Quantitative relevance of substitutional impurities to carrier dynamics in diamond}
\author{Takaaki Shimomura}
\affiliation{Department of Physics, Kyoto University, Kitshirakawa-Oiwake-cho, Sakyo-ku, Kyoto 606-8502, Japan} 
\author{Yoshiki Kubo}
\affiliation{Department of Physics, Kyoto University, Kitshirakawa-Oiwake-cho, Sakyo-ku, Kyoto 606-8502, Japan} 
\author{Julien Barjon}
\affiliation{Groupe d'Etude de la Mati\`ere Condens\'ee (GEMaC), Universit\'e de Versailles St-Quentin-en-Yvelines,
CNRS, Universit\'e Paris Saclay, 45 avenue des Etats-Unis, 78035 Versailles cedex, France}
\author{Norio Tokuda}
\affiliation{Kanazawa University, Kakuma-machi, Kanazawa 920-1192, Japan}
\author{Ikuko Akimoto}
\affiliation{Department of Materials Science and Chemistry, Wakayama University, Wakayama 640-8510, Japan}
\author{Nobuko Naka}\email{naka@scphys.kyoto-u.ac.jp}
\affiliation{Department of Physics, Kyoto University, Kitshirakawa-Oiwake-cho, Sakyo-ku, Kyoto 606-8502, Japan} 
\date{\today}
%\draft

\begin{abstract}
We have quantified substitutional impurity concentrations in synthetic diamond crystals down to sub parts-per-billion levels. The capture lifetimes of electrons and excitons injected by photoexcitation were compared for several samples with different impurity concentrations. 
Based on the assessed impurity concentrations, we have determined the 
capture cross section of electrons to boron impurity, $\sigma_A=1.3\times 10^{-14}$ cm$^{2}$, and that of excitons to nitrogen impurity, $\sigma_D^{ex}=3.1\times 10^{-14}$ cm$^{2}$.
The general tendency of the mobility values for different carrier species is successfully 
reproduced by including carrier scattering by impurities and by excitons.
\end{abstract}

%72.20.Jv Charge carriers: generation, recombination, lifetime, and trapping 
%71.35.Gg Exciton-mediated interactions
%72.10.Fk Scattering by point defects, dislocations, surfaces, and other imperfections (including Kondo effect)  
%72.20.Dp General theory, scattering mechanisms 
%81.05.ug Diamond 

\maketitle

\section{Introduction}

Diamond is a wide-gap material attractive for various usages because of 
its unique physical and chemical properties.
Its practical applications range from ultraviolet light-emitting diodes, radiation detectors, single-photon sources, to biosensors.
Along with the recent progress of crystal growth techniques, such as 
the chemical-vapor-deposition (CVD) and high-pressure high-temperature (HPHT)
methods, high-purity single crystals of diamond are becoming commercially available \cite{review}. It has been known that the dominant substitutional impurities in synthetic diamond are nitrogen and boron due to the proximity of the atomic size to that of carbon. The nitrogen impurities in diamond affect not only the spin relaxation time of electrons localized at nitrogen-vacancy centers \cite{Teraji}, but also relaxation dynamics of free excitons \cite{Naka09}. 

To quantify nitrogen or boron impurities in diamond at parts-per-million (ppm) levels,
one can utilize absorption \cite{Sumiya} and secondary-ion mass spectroscopy (SIMS). However,
methods for evaluating nitrogen at parts-per-billion (ppb) levels below the 
SIMS detection limits are not common \cite{Teraji}.
On the other hand, 
the quantification method for boron impurities by cathodoluminescence \cite{Kawarada}
has been recently extended with the detection limit below 0.05 ppb \cite{BarjonB,BarjonPSSA}.
Because of the long-term difficulty of doping control and the quantification problem,
capture cross sections of carriers or excitons at impurities in diamond have been missing parameters for 20 years \cite{THz, pico}.

In this study, we have performed electron-paramagnetic-resonance measurements to 
determine concentrations of substitutional nitrogen down to sub-ppb levels. 
Also, boron impurities with concentrations below 2 ppb were evaluated by careful photoluminescence analysis. 
On the basis of the assessed impurity concentrations, we quantitatively discuss 
%capture cross sections of charge carriers or excitons and 
momentum relaxation times of charge carriers, which determine the carrier mobility,
in addition to capture cross sections of charge carriers and excitons.
The extraction of the cross sections was enabled by our careful quantification of impurity concentrations.
The capture lifetimes were found to be limited by interactions with impurities in high-purity diamond.

\section{Experiments}

\begin{table}[t]
\caption{List of the samples used in the experiments.
The EPR signal intensity $I$, sample weight $W$, cavity quality factor
$Q$, square root of the microwave power $P$, and the concentration, [N$_s^0$], of 
charge-neutral substitutional nitrogen are given.}
\arraycolsep=5pt
$$
\begin{array}{|c||c|c|c|c|c|} \hline 
\mbox{sample}&I&W \mbox{(mg)}&Q&1/\sqrt{P}&\mbox{[N$_s^0$] (ppb)} \\ \hline 
\mbox{C6}&0.0277&32.1&4700&0.0013&0.07\pm 0.02\\
\mbox{H7}&0.121&4.22&6100&0.0079&2.7\pm 0.5\\
\mbox{C1}&1.37&10.5&5100&0.083&14\pm 1\\
\mbox{H6}&18.2&15.6&5500&0.89& 94\pm 7\\
\mbox{H5}&73.4&18.0&7600&2.2&240\pm 20\\ \hline
\mbox{H2}&429&3.89&6800&83&(23\pm 2)\times 10^3\\ \hline
\end{array}
$$
\label{tab1}
\end{table}

\begin{table}[t]
\caption{Photoluminescence intensity ratio, $R$, of free and boron-bound excitons
at the effective exciton temperature of 12 K
and the boron concentration [B].
$N_i=2$ min([B], [N$_s^0$]) is the expected concentration of ionized impurities due to compensation at thermodynamic equilibrium. H7 is a single-sector crystal whereas H6 and H5 consist of multiple sectors. 
$^*$The value is for the position of PL decay measurement. %113 sector 
%$^{**}$in (111) sectors.
}
$$
\begin{array}{|c||c|c|c|c|} \hline 
\mbox{sample}&\mbox{$R$} &\mbox{[B] (ppb)} &N_i (\mbox{cm}^{-3})\\ \hline 
\mbox{C6}&<0.0005&<0.13& <2.5\times 10^{13}\\ 
\mbox{H7}&0.0062&1.5& 5.3\times10^{14}\\
\mbox{C1}&<0.0015&< 0.4&<1.4\times10^{14} \\
\mbox{H6}^*&0.027&7.2&2.5\times10^{15} \\
\mbox{H5}&0.031\pm0.003 &8.4\pm 1&3.0\times10^{15} \\ \hline
\mbox{C0}&0.38&102&\\ \hline
\end{array}
$$
\label{tab2}
\end{table}

Five samples with unknown impurity concentrations were used in the experiments, 
along with two standard samples (see Table \ref{tab1} for nitrogen and Table \ref{tab2}
for boron quantification).
The first letter of the sample names (C or H) designates the growth method, i.e.,
CVD or HPHT.
The standard sample (H2) for nitrogen quantification was cross-evaluated by 
infrared absorption and SIMS, and the concentration of charge-neutral substitutional nitrogen was [N$^0_s$] $= 23\pm2$ ppm ($=4.1\times 10^{18}$ cm$^{-3}$).
The other standard sample (C0) for boron quantification is an epitaxial layer 
of $13$ $\mu$m-thickness grown on a type Ib diamond substrate.
The boron concentration in the epitaxial layer was characterized by electrical measurements and 
SIMS along the depth \cite{Hall}. The boron concentration averaged over the layer thickness was 
$[B]=$102 ppb $(=1.8\times 10^{16}$ cm$^{-3}$). 
For the samples examined (C6-H5),
the highest possible concentrations quoted by the suppliers were 50 ppb for boron 
and 100 ppb for nitrogen \cite{Horiuchi}.

A substitutional nitrogen in diamond forms a nitrogen P$_1$ center. 
Two electronic levels arise by the Zeeman effect under an applied magnetic field. Each level is
further split into triplets by the nuclear spin interaction originating from $^{14}$N
atoms with natural abundance of 99.64 $\%$.
The concentration of unpaired electrons, or that of the P$_1$ centers
can be determined based on the microwave absorption associated with the transitions 
between these electronic levels \cite{Wyk}. 
The electron paramagnetic resonance (EPR) signal was measured with 
X-band microwave at a frequency of 9.6 GHz
for samples mounted in a dielectric cavity (Bruker, MD-5-W1, TE011) at room temperature. 
The measurement was done in 
the continuous-wave regime by using the spectrometer (Bruker, ELEXSYS E580).
The nitrogen concentration was obtained by using the relation:
\begin{equation}
[N_s^0]\propto I/(WQ\sqrt{P}),
\end{equation}
where $I, W, Q,$ and $P$ stand for
the EPR signal intensity, weight of the sample, cavity quality factor, and microwave power,
respectively. 
The EPR signal intensity was evaluated at the limit of weak microwave power
to avoid the saturation effect.
This was facilitated by analyzing the power dependence of the EPR signals, 
and then by choosing appropriate microwave power from sample to sample for quantification
(see Table \ref{tab1}).
The absolute concentration was evaluated by using the signal intensity from the standard sample (H2).

The boron concentration [B] was determined based on the photoluminescence 
intensity of boron-bound excitons relative to that of free excitons \cite{BarjonB}. 
The excitation source was 213 nm wavelength light from a tunable laser based on 
an optical-parametric-oscillator system (Ekspla, NT242). 
The pulse duration was 2.5 ns and the repetition rate was 1 kHz. 
Instead of the continuous excitation used in Ref. \cite{BarjonB}, pulsed excitation was used here but with a minimized laser power (e.g., 0.4 $\mu$W), so as to achieve suitable conditions for the quantification.
The measurements were performed at 7 K for samples mounted in a closed-cycle cryostat. 
%, or 2 K for samples immersed in superfluid helium. 
The photoluminescence signal was detected by a CCD camera (Andor, DU940N-BU2)
at the exit of a monochromator (Horiba Jobin Yvon, iHR-550, 2400 grooves/mm grating).
The entrance slit of the monochromator was set at 150 $\mu$m
to achieve sufficient signal intensities under weak excitation. 
The absolute boron concentration was determined by comparing the 
PL intensity ratio with that of the standard sample (C0) measured under identical conditions.

\section{Results and discussion}
\subsection{Impurity quantification}

\begin{figure}[tb]
\centering
\includegraphics[width=8.5cm,clip]{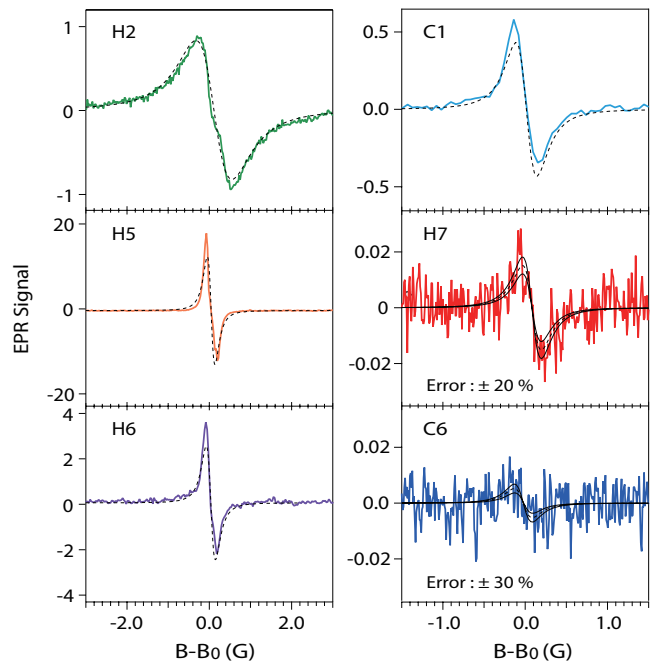}
\caption{EPR spectra of the (-1/2, 0) $\rightarrow$ (1/2, 0) transition 
of the substitutional nitrogen (P$_1$) center in various samples.
Dashed lines are the best fit functions. Thin full lines are drawn to estimate the errors
($\epsilon=\pm 0.2$ for H7 and $\epsilon=\pm 0.3$ for C6).}
\label{fig:01}
\end{figure}

Figure \ref{fig:01} shows EPR spectra obtained from each sample. 
Here, we focus on the (-1/2, 0) $\rightarrow$ (1/2, 0) transition, 
where the numbers in parentheses indicate the electron spin and the nuclear spin. 
The center magnetic field $B_0$ corresponds to $g=2.0027 \pm 0.0002$.
The spectral shape of the EPR signals is generally given by a Lorentzian function, but here it is represented in a differential form due to the magnetic field modulation (modulation width = 0.1 Gauss) to improve the sensitivity.
The EPR signal intensity $I$ in Eq. (1) is obtained by integrating a raw signal twice over the magnetic field strength.
Alternatively, $I=\pi ab$ is obtained by fitting the data with a differentiated spectral function,
\begin{equation}
L'(B)=-ab^2\frac{B-B_0}{[b^2+(B-B_0)^2]^2}.
\end{equation}
The dashed lines represent the best fit functions. The evaluated signal intensity $I$
and estimated nitrogen concentration [N$_s^0$] are summarized in Table \ref{tab1}.
The errors for a concentration below the ppb level were 
carefully evaluated by comparing the data with the 
 best fit function magnified by a factor of $(1+\epsilon)$, where $\epsilon$ yields 
the relative uncertainty. 
These spectral functions are shown by thin full lines in Fig. 1.
The minimum nitrogen concentration estimated was of the order of 0.1 ppb,
which corresponds to $1.76\times 10^{13}$ cm$^{-3}$.

Concerning the boron concentration, we referred to the previous publication \cite{BarjonB} on the ratio $R$
%=I_{BE}/I_{FE}$ 
of cathodoluminescence from bound and free excitons 
in boron-doped diamond. In order to reduce errors during the quantification procedure, 
we mounted three samples in the same holder to compare.
The obtained photoluminescence intensity ratio and the boron concentration are summarized in Table \ref{tab2}.
The photoluminescence from bound excitons was below the detection limit in C1 and C6, thus we 
provided in Table \ref{tab2} an upper bound for the boron concentration by considering the signal-to-noise ratio.

\subsection{Capture lifetimes}

\begin{figure}[t]
\centering
\includegraphics[width=7.5cm,clip]{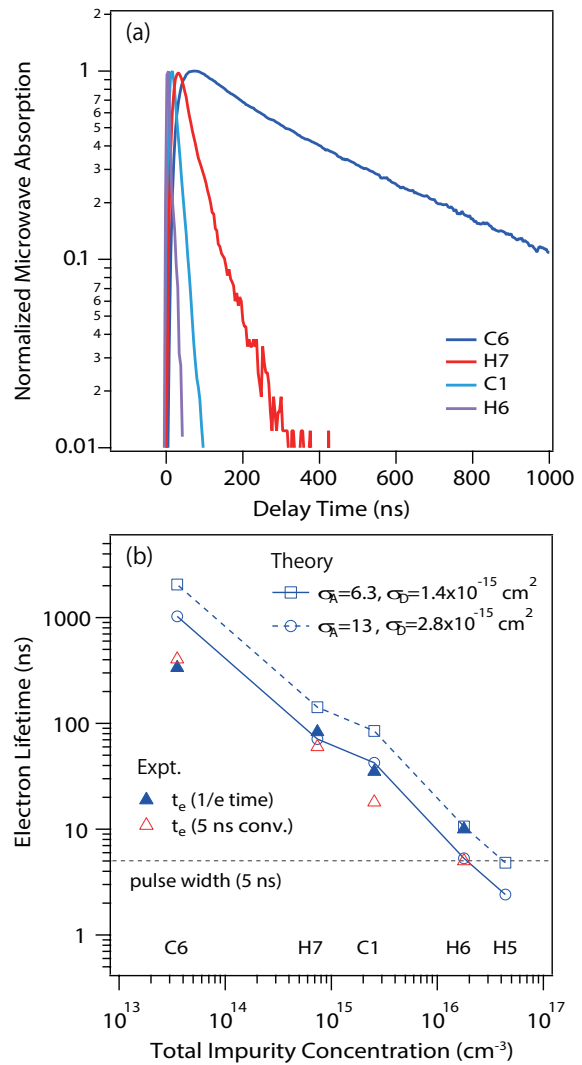}
\caption{(a) Normalized intensity of the cyclotron resonance signal for electrons 
at 10 K in various samples. (b) Plot of capture lifetime as a function of total impurity concentration.}
\label{fig:02}
\end{figure}

The above result indicates that the nitrogen concentration is higher than the boron concentration in most samples.
%(H5, H6, C1, H7) except for the highest-purity one (C6). 
By using these assessed samples having different impurity concentrations, we discuss influence of impurities on the exciton or carrier capture lifetimes.
Figure \ref{fig:02}a shows normalized intensity of cyclotron resonance signals for electrons 
at 10 K in various samples measured in our previous work with 
photoexcitation at around 223 nm \cite{AkimotoDRM,NakaPSSA}. 
The decay time was extracted as a 1/e time or by convolving an exponential
decay function with a Gaussian responsible for the excitation pulse and the system response function.
The decay time largely depended on samples.  
Figure \ref{fig:02}b is a plot of measured decay time as a function of the total 
impurity concentration. 
That is, the horizontal axis of the figure is [$N_s^0$]+[B].
The symbols connected with lines are capture lifetimes calculated by
\begin{equation}
t=\frac{1}{v(\sigma_A n_A +\sigma_D n_D)},
\end{equation}
where, $v=\sqrt{3k_BT/m^*_{dos}}$ ($m^*_{dos}=0.496 m_0$ \cite{Naka13}) 
is the electron thermal velocity determined by the temperature $T$,
$\sigma_A, \sigma_D$ are the cross sections of electrons against capture to boron and nitrogen, and
$n_A, n_D$ are acceptor or donor concentrations. We take $n_A=[B]$ and $n_D=[N_s^0]$ 
by assuming neutralization of compensated impurities by photoexcited carriers as proposed in Refs. \cite{Fukai,Barjon2007}.

The cross section, $\sigma_D=1.4 \times 10^{-15}$ cm$^2$, for nitrogen at 10 K was taken 
from terahertz time-domain spectroscopy \cite{THz} on electron capture at nitrogen.
The cross section for boron was estimated considering the difference of the 
effective Bohr radius of impurity states, as given by $a_A=\hbar/\sqrt{2m_0 E_A}=5.5 \AA$ 
and $a_D=\hbar/\sqrt{2m_0 E_D}=2.6 \AA$,
where $E_{A(D)}= 0.37 (1.7) $ eV is the acceptor (donor) activation energy.
Here, a hydrogen model was assumed both for boron and nitrogen levels.
According to the model of classical collision between hard spheres,
the collision cross section is given by $\sigma_{A(D)}=4 \pi a_{A(D)}^2$.
This scaling relation yields $\sigma_A=\sigma_D a_A^2/a_D^2=6.3 \times 10^{-15}$ cm$^2$, and the calculated capture lifetimes are shown by open squares.
A better agreement with data was obtained with slightly larger cross sections 
$\sigma_A=1.3 \times 10^{-14}$ cm$^2$ and
$\sigma_D=2.8 \times 10^{-15}$ cm$^2$ (circles).
The measured lifetime for C6 was shorter than the calculated ones
assuming the boron concentration at the upper bound. This implies
existence of other kind of structural defects or impurities, 
such as hydrogen, silicon, interstitials, or dislocations, acting as a center capturing electrons in the CVD-grown diamond.
Further, because of the larger cross section for boron than for nitrogen,
the electron capture time more sensitively depends on the boron concentration.
This is why the plot does not look linear to the total impurity concentration. 

\begin{figure}[t]
\centering
\includegraphics[width=7.5cm,clip]{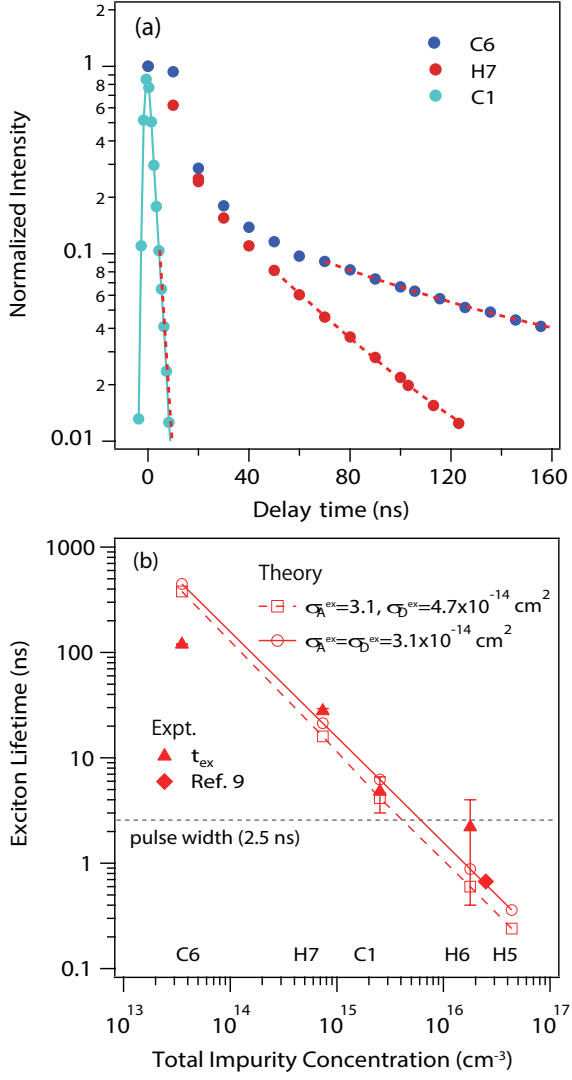}
\caption{(a) Normalized PL intensity of free-exciton luminescence in C1, H7 and C6 at 7 K. 
(b) Plot of capture time (lifetime) as a function of total impurity concentration.}
\label{fig:03}
\end{figure}

Figure \ref{fig:03}a shows normalized PL intensity of free-exciton luminescence following the 2.5 ns pulsed excitation in C1, H7 and C6 at 7 K. Detailed experimental setup for this measurement is 
described elsewhere \cite{Morimoto}. After the fast decay due to many-body effects, 
the 1/e time of the second decay component indicates the time for free excitons 
to be captured by impurities. The capture lifetimes were obtained as 4.8, 28, and 430 ns 
for C1, H7, and C6, respectively. 
The analysis for decay times shorter than the duration of the laser pulse involves large errors because separating many-body effects is not straightforward. Nevertheless, there is a clear tendency that
the capture lifetime was shorter for a sample with a higher concentration of impurities. 
Figure \ref{fig:03}b is a plot of the capture lifetime as a function of 
total impurity concentration.
The calculations were performed based on Eq. (3),
where $v=\sqrt{3k_BT/m^*_{ex}}$ is the exciton thermal velocity 
with the translational effective mass $m_{ex}^*=0.76 m_0$ \cite{Naka13}
%m_{c}^*+m_{so}^*=
 and $\sigma_A, \sigma_D$ should be replaced by the cross sections of excitons, 
$\sigma_A^{ex}$ and $\sigma_D^{ex}$.
We assumed $\sigma_A^{ex}=3.1 \times 10^{-14}$ cm$^2$ after Ref. \cite{pico} for boron, where 
a factor of $\sqrt{8/(3\pi)}$ was multiplied to correct the difference in definitions of 
thermal velocity. The cross section for nitrogen is not exactly known, 
because the direct measurement is difficult due to the fast Auger process.
For the calculation as shown by squares, the capture cross section 
$\sigma_D^{ex}=4.7 \times 10^{-14}$ cm$^2$ for phosphorous \cite{pico} 
is used in place of that for nitrogen.
A reasonable agreement with the data was obtained when we adjusted the 
cross section for nitrogen as $\sigma_D^{ex}=3.1\times 10^{-14}$ cm$^2$, as shown by circles. 

Here we would like to summarize the obtained cross sections
and compare the values for carriers and for excitons.
The electron cross section for boron  
$\sigma_A=1.3 \times 10^{-14}$ cm$^2$ is larger than that for nitrogen
$\sigma_D=2.8 \times 10^{-15}$ cm$^2$,
and both are smaller than the exciton cross sections 
$\sigma_D^{ex}=\sigma_A^{ex}=3.1 \times 10^{-14}$ cm$^2$.
This means that electrons have less chances to be captured by impurities than excitons. 
The difference seems to come from the difference in the Bohr radius ($a_A$, $a_D$) of carriers 
and that ($a_B^{ex}$) of excitons. Again, by assuming classical collision between hard spheres,
the Bohr radius scales proportionally to $\sqrt{\sigma}$. 
Therefore, the excitonic Bohr radius is roughly estimated as $a_B^{ex}=a_A\sqrt{\sigma_A^{ex}/\sigma_A}$
 by using the acceptor cross sections, and $a_B^{ex}=a_D\sqrt{\sigma_D^{ex}/\sigma_D}$ by using the donor cross sections. Both relations coherently lead to an exciton Bohr radius of 8.6 $\AA$.
Considering the crudeness of the hydrogen model, this value is not far from 
the value, $a_B^{ex}=13.7$ $\AA$, derived from the exciton binding energy of 92 meV\cite{Sauer}.
Therefore, exciton capture is determined by the exciton Bohr radius rather than 
by the effective radius of impurities.

The good agreement obtained above between the experimental data and theory
implies that neglecting ionized impurities is reasonable. 
Additionally, capture by impurity complexes such as nitrogen-vacancy centers should not be 
significant because
the density of the centers are small, at least 1/200 times [N$_s^0$] \cite{Teraji2},
and the cross section is known to be very small ($10^{-15}$ cm$^2$) \cite{NVcapture}.
The excitons in diamond are bound by neutral impurities but unbound by charged impurities due to the relatively small mass imbalance between electrons and holes \cite{Dean}. 

\subsection{Momentum relaxation time of charge carriers}

\begin{figure*}[bht]
\centering
\includegraphics[width=17cm,clip]{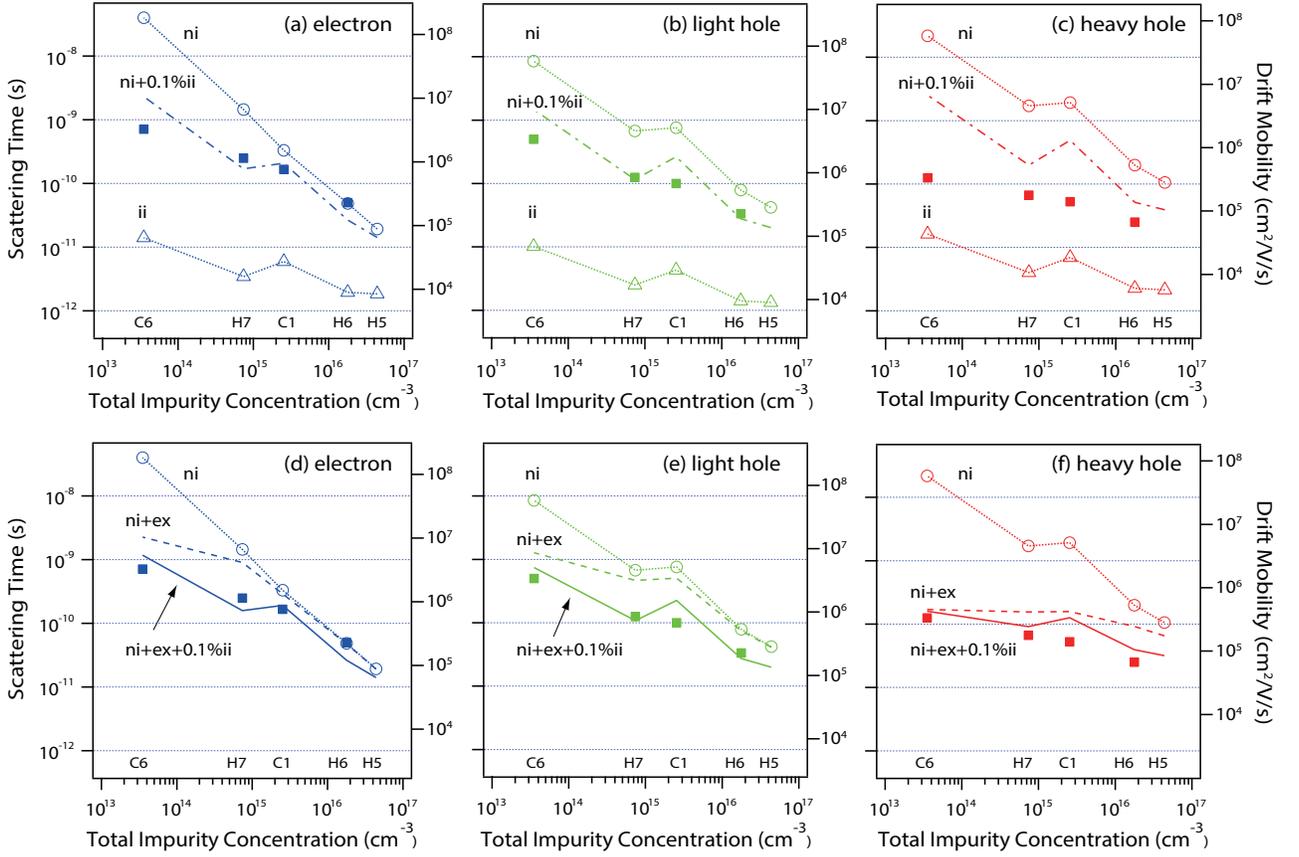}
\caption{Solid squares: scattering times measured at 10 K in different samples for 
(a,d) electron, (b,e) light hole, and (c,f) heavy hole, respectively.
Left axes and grids are common for all six panels.
Symbols connected by lines represent calculations of neutral impurity (ni) and ionized impurity (ii)
scattering. The dashed dotted lines in (a,b,c) represent 
the sum of (ni) and (ii) scattering calculated with $0.1\%$ $N_i$.
The dashed lines in (d,e,f) are the sum of (ni) and exciton (ex) scattering, and the full lines are the 
sum of (ni), (ex) and (ii) scattering with 0.1$\%$ $N_i$. 
%%%%%%%%%%%%%%%%%%%%%%%%%%{\color{green}(add error bars)}
}
\label{fig:04}
\end{figure*}

In this subsection, we compare the momentum relaxation time (or the scattering time in short)
of charge carriers in different samples.
The thermal average of the scattering time, $\tau$,
 is directly related to drift mobility by the relation $\mu=e\tau/m_c^*$,
where $e$ is the elementary charge and $m_c^*$ is the conductivity effective mass. 
According to the Mathiessen's rule, the scattering time is approximated by 
the inverse of the sum of the rates due to intrinsic and extrinsic scattering processes. 
The intrinsic process is well known to be dominated by phonon scattering \cite{AkimotoAPL}.
If the extrinsic scattering processes are dominated by impurity interactions, one can, in principle, predict the drift mobility
of charge carriers in assessed samples, after precisely knowing impurity concentrations.

The measurements of the scattering time were done by means of cyclotron resonance \cite{NakaPSSA}, where the ratio of the resonance width to the resonance magnetic field yields the value of $\tau$.
With decreasing temperature $T$, the scattering time $\tau$ increased as $T^{-3/2}$ 
down to 50 K, in agreement with the acoustic-phonon scattering theory. 
At lower temperatures, $\tau$ approached a constant value
with a weaker temperature dependence than $T^{-3/2}$.
The low-temperature scattering time was shorter for diamond containing 
impurities of higher concentrations (see the plot of 1/$\tau$ in Fig. 12 of Ref. \cite{NakaPSSA}).

The scattering times at 10 K in assessed samples are plotted by solid squares in Figs. \ref{fig:04}(a)-(c), respectively for the electron, light, and heavy hole.
The longest scattering time was measured for the electron in C6 whereas 
the shortest one was measured for the heavy hole in H6.
The resonance spectra in H5 were too broad to extract an accurate scattering time.
%%%%%%%%%%%%%%%%%%%%%%%%%%%%%%%%%% \textcolor{green}{(see the error bars in Figs. \ref{fig:04})}. 
The right axis of each panel represents the mobility value obtained by using the conductivity effective masses, $m_e^*=0.385 m_0$, $m^*_{lh}=0.26m_0$, and $m^*_{hh}=0.66m_0$.

The most obvious contribution to charge carrier scattering comes from neutral impurities (ni), whose theoretical scattering time is given 
by Erginsoy formula \cite{Erginsoy} for an electron-hydrogen scattering model
and its modification to a positron-hydrogen scattering picture by Otsuka, Murase, and Iseki \cite{Otsuka}. That is,
\begin{equation}
\tau_{ni-e}= m^*_{dos}/(3.4 n_A a_A \hbar+20 n_D a_D\hbar) \propto m_{dos}^*,
\end{equation}
for an electron, and 
\begin{equation}
\tau_{ni-h}= m^*_{hh(lh)}/(20 n_A a_A\hbar+3.4 n_D a_D\hbar) \propto m_{hh(lh)}^*,
\end{equation} 
for a hole.
As shown by circles in Figs. \ref{fig:04}(a)-(c),  
the calculated scattering times for (ni) are quite longer than experimental data,  especially for the heavy hole. In the following paragraphs we discuss the other possible contributions. 

A source for the short scattering times measured for the heavy hole could be fluctuation in isotopic composition of natural carbon (98.9 $\%$ $^{12}$C and 1.1 $\%$ $^{13}$C). 
The isotope scattering is modeled by alloy scattering \cite{alloy} with the scattering time expressed as,  
\begin{eqnarray}
\tau_{iso}=\frac{(2\pi\hbar^2)^2}{\pi m_{dos}^{*2}}
\left(\frac{dE_c}{dx}\right)^{-2} \frac{1}{a_0^3 x \sqrt{3k_BT/m_{dos}^*}} \propto m_{dos}^{*-3/2},
\end{eqnarray}
where $x=0.011$ is the fraction of $^{13}$C, $a_0=3.57 \AA$ is the lattice constant of diamond, and $dE_c/dx=14.6$ meV is the isotopic gap shift \cite{Kanda}.
The calculated scattering times are $\tau_{iso}=$ 1.5$\times10^{-7}$ s, 2.2$\times10^{-7}$ s, and 5.5$\times10^{-8}$ s for the electron, light hole, and heavy hole, respectively.
As these values are much larger than the experimental data, we conclude that their contributions to the mobility are not significant.

A local strain could also be a source for carrier scattering. 
An order-of-magnitude scattering time is in the $10^{-10}$ s range according to Eq. (3) of Ref. \cite{disloc}:
\begin{equation}
\tau_{strain}=\frac{32}{3\pi}\left(\frac{1-\nu}{1-2\nu} \right)^2\frac{k_B\hbar}{\Xi^2
\lambda^2N}T,
\end{equation}
where $\nu=0.18$ is Poisson ratio, $\Xi=8.7 (10)$ eV is the deformation potential for 
the electron (light and heavy holes) \cite{AkimotoAPL}, and $\lambda$ is 
the unit crystallographic slip distance taken to be equal to the lattice constant.
Here we assumed dilations around edge-type dislocations at a density of $N=10^5$ cm$^{-2}$, 
which is a typical value in samples with the same grade as C6 and orders of magnitude greater than that in HPHT samples (H7 is dislocation-free). 
The theoretical scattering time is independent of the effective mass,
and does not explain the discrepancy between (ni) and squares in Fig. \ref{fig:04}.

On the other hand, the case of ionized impurities (ii) deserves to be considered since samples are compensated. Calculated scattering times for ionized impurity interaction according to the Conwell Weisskopf formula \cite{Ridley}, 
\begin{equation}
\tau_{ii}=\frac{64\sqrt{\pi m_{dos}^*}\epsilon^2(2k_BT)^{3/2}}{e^4 N_i\ln [(1+144\pi^2\epsilon^2k_B^2T^2/(e^4N_i^{2/3})]} \propto m_{dos}^{*1/2},
\end{equation}
are represented by triangles in Figs. 4(a)-(c) when using the density $N_i=2$ min([B], [N$_s^0$]) as listed in Table II
\footnote{Given a 10$^{4-5}$ cm$^{-2}$ dislocation density, it would provide
a negligible amount of charged centers compared to $N_i$.}
and the dielectric constant $\epsilon=5.7$.
With such inputs, the calculated (ii) scattering times are about two orders of magnitudes shorter than the measured ones. 
This result indicates that the actual density of ionized impurities is much smaller than $N_i$. 
The dashed dotted lines in Figs. 4(a)-(c) are instead calculated with the neutral impurities and a concentration of ionized impurities equal to 0.1$\%$ $N_i$. The relatively good agreement with experimental data indicates that at least 99.9$\%$ of the compensated impurities are neutralized by photoexcited carriers, 
in consistency with the assumption made in Subsection B for the analysis of capture lifetimes. 
Our result also broadens the context known by the first experiments in silicon \cite{Fukai},
where ionized impurity scattering plays a minor role in cyclotron resonance under photoexcitation
compared to Hall-effect measurements.

Finally, we had to consider scattering between free carriers and excitons \cite{Ohyama}
because the short scattering times for the heavy hole are still not correctly described 
with both neutral and ionized impurities. The carrier-exciton collision
 occurs with a scattering time $\tau_{ex}=1/(Q n_{ex} v),$ where $Q$ is the 
exciton-carrier cross section.
The cross section depends on the mass ratio of scattering particles 
and thus on the carrier species. Our estimates based 
on the literature \cite{Elkomoss} give rise to 
$Q_e=2.6\pi (a_B^{ex})^2=1.4\times 10^{-13}$ cm$^2$,
$Q_{lh}=3\pi (a_B^{ex})^2=1.6\times 10^{-13}$ cm$^2$, and
$Q_{hh}=42\pi (a_B^{ex})^2=2.2\times 10^{-12}$ cm$^2$.
The exciton-carrier cross section for the heavy hole is found to be 
one order of magnitude higher than that for the electron or light hole.
This can be understood by the fact that
the interaction potential for exchange of the incident carrier and that of 
the exciton becomes repulsive when the mass ratio exceeds unity, which is the only case 
for the heavy hole.
The calculated scattering time for the exciton density of $n_{ex}=
10^{15}$ cm$^{-3}$ is $2.4\times 10^{-9}$ s, $1.5\times 10^{-9}$ s, and $1.7\times 10^{-10}$ s
for the electron, light hole, and heavy hole, respectively at $T=10$ K.  
The particular behavior of heavy-hole scattering times as measured by cyclotron resonance is probably a signature of a significant exciton-carrier scattering. At the end,
the dashed lines in Figs. 4(d)-(f) correspond to 
the sum of (ni) and exciton scattering. By adding the 
contributions of 0.1$\%$ $N_i$ ionized impurity scattering, as shown by the full lines,
a reasonable agreement with the data (solid squares) is obtained.

Although we have not reached a perfect agreement, our first attempt to preliminary determine 
the mobility seems to be successful to reproduce the general tendency.
It is unique to diamond that both charge carriers and excitons coexist up to room temperatures and above.
Since such a situation is realized only at cryogenic temperatures in prototypical semiconductors like Si and GaAs, further investigation of diamond at higher temperatures would be interesting. On the other hand, 
studies toward the other extreme, i.e., higher purity diamond at ultralow temperatures
in the absence of neutral impurity scattering would also be fascinating. 
When a higher sensitivity for detection is achieved and carrier-exciton scattering is suppressed, carrier mobility limited by isotope scattering might be seen for the first time in semiconductors.

\section{Conclusion}
We have quantified substitutional nitrogen and boron concentrations in 
several synthetic diamond crystals down to sub ppb levels.  
The capture lifetimes of electrons and excitons as well as the carrier scattering times
were discussed as a function of impurity concentrations.
We extracted the cross section of electrons by boron impurity, and that of excitons 
by nitrogen impurity, which had been unknown so far for diamond.
The scattering times of charge carriers with impurities, carbon isotopes, and excitons were quantitatively 
calculated, and the comparison with measured data 
indicated an almost complete neutralization of compensated impurities by photoexcited carriers.

The present study opens a new avenue to access the detailed understanding of dynamics 
of the coexistent system of charge carriers and excitons. 
The information obtained in the present study should be useful for predicting mobility-lifetime ($\mu\tau$) products in intrinsic layers of diamond diodes 
as well as charge carrier collection efficiencies in diamond detectors.

\section*{Acknowledgements}
This work was partially supported by JSPS
KAKENHI Grants Numbers 17H02910, 15K05129, 26400317, and The Murata Science Foundation.
N. N. thanks CNRS for financial support for an invited researcher position at GEMaC,
Solange Temgoua (GEMaC, Versailles University UVSQ) for cathodoluminescence measurements,
Makoto Kuwata-Gonokami (The University of Tokyo) for courtesy of HPHT samples,
and Yuzo Shinozuka (Wakayama University) for valuable discussion on alloy scattering.


\begin{thebibliography}{99}
	\bibitem{review}
	R. S. Balmer, J. R. Brandon, S. L. Clewes, H. K. Dhillon, J. M. Dodson,
	I. Friel, P. N. Inglis, T. D. Madgwick, M. L. Markham, T. P. Mollart,
	N. Perkins, G. A. Scarsbrook, D. J. Twitchen, A. J. Whitehead,
	J. J. Wilman, and S. M. Woollard,
	Chemical vapour deposition synthetic diamond: materials, technology and applications,
	J. Phys.: Condens. Matter {\bf 21} 364221 (2009).
	\bibitem{Teraji}
	T. Teraji, J. Isoya, K. Watanabe, S. Koizumi, and Y. Koide,
	Homoepitaxial diamond chemical vapor deposition for ultra-light doping,
	Materials Science in Semiconductor Processing {\bf 70} 197 (2017).
	\bibitem{Naka09}
	N. Naka, J. Omachi, H. Sumiya, K. Tamasaku, T. Ishikawa, and M. Kuwata-Gonokami,
	Density-dependent exciton kinetics in synthetic diamond crystals,
	Phys. Rev. B {\bf 80} 035201 (2009).
	\bibitem{Sumiya}
	H. Sumiya and S. Satoh, High-pressure synthesis of high-purity diamond crystal, Diam. Relat. Mater. {\bf 5} 1359 (1996).
	\bibitem{Kawarada}
	H. Kawarada, H. Matsuyama, Y. Yokota, T. Sogi, A. Yamaguchi, and A. Hiraki,
Excitonic recombination radiation in undoped and boron-doped chemical-vapor-deposited diamonds, 	
	Phys. Rev. B {\bf 47} 3633 (1993).
	\bibitem{BarjonB}
	J. Barjon, T. Tillocher, N. Habka, O. Brinza, J. Achard, R. Issaoui, F. Silva,
	C. Mer, and P. Bergonzo, 
	Boron acceptor concentration in diamond from excitonic recombination intensities,
	Phys. Rev. B {\bf 83} 073201 (2011).
	\bibitem{BarjonPSSA} J. Barjon, Luminescence Spectroscopy of Bound Excitons in Diamond, Phys. Status Solidi A {\bf 214} 1700402 (2017).
	\bibitem{THz}
	R. Ulbricht, S. T. van der Post, J. P. Goss, P. R. Briddon, R. Jones, R. U. A. Khan, M. Bonn,
	Single substitutional nitrogen defects revealed as electron acceptor states in diamond using ultrafast spectroscopy,
	Phys. Rev. B {\bf 84} 165202 (2011).
	\bibitem{pico}
	J. Barjon, P. Valvin, C. Brimont, P. Lefebvre, O. Brinza, A. Tallaire, J. Achard, 	F. Jomard, and M. A. Pinault-Thaury,
	Picosecond dynamics of free and bound excitons in doped diamond,
	Phys. Rev. B {\bf 93} 115202 (2016).
	\bibitem{Hall}
	J. Barjon, E. Chikoidze, F. Jomard, Y. Dumont, M.-A. Pinault-Thaury,
	R. Issaoui, O. Brinza, J. Achard, and F. Silva,
	Homoepitaxial boron-doped diamond with very low compensation, Phys. Status Solidi A {\bf 209} 1750 (2012).
	\bibitem{Horiuchi}
	K. Horiuchi, K. Nakamura, S. Yamashita, and M. Kuwata-Gonokami,
	Photoluminescence characterization of high-purity synthesized diamond,
	Jpn. J. Appl. Phys. {\bf 36} L1505 (1997).
	\bibitem{Wyk}
	J. A. van Wyk, E. C. Reynhardt, G. L. High, and I Kiflawi,
	The dependences of ESR line widths and spin - spin relaxation times of single nitrogen defects on the concentration of nitrogen defects in diamond,
	J. Phys. D: Appl. Phys. {\bf 30} 1790 (1997).
	\bibitem{NakaPSSA}
	N. Naka, H. Morimoto, and I. Akimoto,
	Excitons and fundamental transport properties of diamond under photo-injection,
	Phys. Status Solidi A {\bf 213} 2551 (2016).
	\bibitem{AkimotoDRM}
	I. Akimoto, N. Naka, and N. Tokuda,
	Time-resolved cyclotron resonance on dislocation-free HPHT diamond,
	Diam. Relat. Mater. {\bf 63} 38 (2016).
	\bibitem{Naka13} 
	N. Naka, K. Fukai, Y. Handa, and I. Akimoto,
	Direct measurement via cyclotron resonance of the carrier effective masses in pristine diamond,
	Phys. Rev. B {\bf 88} 035205 (2013).
	\bibitem{Fukai} 
	M. Fukai, H. Kawamura, K. Sekido, and I. Imai,
	Line-Broadening of Cyclotron Resonance due to Lattice and Neutral Impurity Scattering in Silicon and Germanium,
	J. Phys. Soc. Jpn {\bf 19} 30 (1964).
	\bibitem{Barjon2007}
	J. Barjon, M.-A. Pinault, T. Kociniewski, F. Jomard, and J. Chevallier,
	Cathodoluminescence as a tool to determine the phosphorus concentration in diamond,
	Phys. Status Solidi A {\bf 204} 2965 (2007).
	\bibitem{Morimoto}
	H. Morimoto, Y. Hazama, K. Tanaka, and N. Naka,
	Exciton lifetime and diffusion length in high-purity chemical-vapor-deposition diamond,
	Diam. Relat. Mater. {\bf 63} 47 (2016).
	\bibitem{Sauer}
	R. Sauer, N. Teofilov, and K. Thonke, Exciton condensation in diamond, Diam. Relat. Mater. {\bf13} 691 (2004).
	\bibitem{Teraji2}
	T. Teraji, T. Yamamoto, K. Watanabe, Y. Koide, J. Isoya, S. Onoda, T. Ohshima,
	L.J. Rogers, F. Jelezko, P. Neumann, J. Wrachtrup, and S. Koizumi,
	Homoepitaxial diamond film growth: High purity, high crystalline quality, isotopic enrichment, and single color center formation,
	%Homoepitaxial diamond film growth: High purity, high crystalline quality, isotopic enrichment,
	%and single color center formation, 
	Phys. Status Solidi A {\bf 212} 2365 (2015).
	\bibitem{NVcapture}
	D. Yu Fedyanin and M. Agio,
	Ultrabright single-photon source on diamond with electrical pumping at room and high temperatures,
	New J. Phys. {\bf 18} 073012 (2016).
	\bibitem{Dean}
	P. J. Dean and D. C. Herbert, {\it Excitons} eds. by K. Cho (Springer, 1979).
	\bibitem{AkimotoAPL}
	I. Akimoto, Y. Handa, K. Fukai, and N. Naka,
	High carrier mobility in ultrapure diamond measured by time-resolved cyclotron resonance,
	Appl. Phys. Lett. {\bf 105} 032102 (2014).
	%\bibitem{12C}
	%H. Watanabe and C. E. Nebel, Diam. Relat. Mater. {\bf 17} 511 (2008).
	\bibitem{Erginsoy} C. Erginsoy, Neutral Impurity Scattering in Semiconductors, Phys. Rev. {\bf 79} 1013 (1950).
	\bibitem{Otsuka}
	E. Otsuka, K. Murase, and J. Iseki,
	Electron Scattering by Neutralized Acceptors in Germanium I. Gallium and Indium,
	J. Phys. Soc. Jpn. {\bf 21} 1104 (1966).
	\bibitem{alloy}
	S. Fahy, and E. P. O'Reilly,
	Intrinsic limits on electron mobility in dilute nitride semiconductors,
	App. Phys. Lett. {\bf 83} 3731 (2003).
	\bibitem{Kanda}
	T. Ruf, M. Cardona, H. Sternschulte, S. Wahl, K. Thonke, R. Sauer, P. Pavone and T.R. Anthony,
	Cathodoluminescence investigation of isotope effects in diamond,
	Solid State Commun. {\bf 105} 311 (1998).
	\bibitem{disloc}
	%B. Podor, Phys. Status Solidi {\bf 16} K167 (1966).
	%M. Nagabhooshanam and V. Hari Babu, J. Mat. Science, {\bf 20} 4329 (1985).
	D. L. Dexter and F. Seitz, Effects of Dislocations on Mobilities in Semiconductors, Phys. Rev. {\bf 86} 964 (1952).
	\bibitem{Ridley}
	B. K. Ridley, {\it Quantum Processes in Semiconductors} (Oxford).
	\bibitem{Ohyama}
	T. Ohyama, T. Sanada, and E. Otsuka,
	Time-Resolved Cyclotron Resonance Analysis of Electron-Exciton Interaction in Silicon,
	J. Phys. Soc. Jpn {\bf 35} 822 (1973).
	\bibitem{Elkomoss}
	S. G. Elkomoss and G. Munschy,
	Electron-exciton elastic scattering cross sections in the central field and the exchange approximations,
	J. Phys. Chem. Solids {\bf 38} 557 (1977).

\end{thebibliography}
\end{document}